\documentclass[fleqn,10pt]{SelfArx} 

\usepackage{lipsum} 

\definecolor{color1}{RGB}{0,0,90} 
\definecolor{color2}{RGB}{0,20,20} 

\usepackage{hyperref} 
\hypersetup{hidelinks,colorlinks,breaklinks=true,urlcolor=color2,citecolor=color1,linkcolor=color1,bookmarksopen=false,pdftitle={Title},pdfauthor={Author}}

\JournalInfo{Accepted to Astronomy Reports, 2015, Vol. 59, No. 7} 
\Archive{}

\PaperTitle{
Solar-Type Activity: Epochs of Cycle Formation
} 

\Authors{M. M. Katsova\textsuperscript{1}*, N. I. Bondar'\textsuperscript{2}, M. A. Livshits\textsuperscript{3}} 
\affiliation{\textsuperscript{1}\textit{
Lomonosov Moscow State University, Sternberg Astronomical Institute,
Moscow, Russia}} 
\affiliation{\textsuperscript{2}\textit{
Institute of Astronomy, Russian Academy of Sciences, Moscow, Russia
}} 
\affiliation{\textsuperscript{3}\textit{
Pushkov Institute of Terrestrial Magnetism, Ionosphere, and Radio-Wave Propagation,\\
 ~~ Russian Academy of Sciences, Moscow, Troitsk, Russia
}} 
\affiliation{*\textbf{Corresponding author}: maria@sai.msu.ru} 

\newcommand{\keywordname}{Keywords} 

\Abstract{
The diagram of indices of coronal and chromospheric activity allowed us to
reveal stars where solar-type activity appears and regular cycles are
forming. Using new consideration of a relation between coronal activity and
the rotation rate, together with new data on the ages of open clusters, we
estimate the age of the young Sun corresponding to the epoch of formation
of its cycle. The properties of the activity of this young Sun, with an age
slightly older than one billion years, are briefly discussed. An analysis
of available data on the long-term regular variability of late-type stars
leads to the conclusion that duration of a cycle associated with solar-type
activity increases with the deceleration of the stellar rotation; i.e.,
with age. New data on the magnetic fields of comparatively young G stars
and changes in the role of the large-scale and the local magnetic fields in
the formation of the activity of the young Sun are discussed. Studies in
this area aim to provide observational tests aimed at identifying the
conditions for the formation of cyclic activity on stars in the lower part
of the main sequence, and test some results of dynamo theory.
}

\begin{document}

\flushbottom 

\maketitle 

\tableofcontents 

\thispagestyle{empty} 


\section*{Introduction} 

\addcontentsline{toc}{section}{Introduction} 

There is now a large number of observations of
individual solar phenomena demonstrating various
specific features of the present solar activity. Two
basic methods can be used to understand the solar
activity in the past, during the epoch of formation of
the solar cycle. The first deals with the evolution of
the angular momentum of the stellar axial rotation
and dynamo theory. This provides the possibility of
identifying the main factor of the  evolution of activity,
related to the deceleration of the rotation. The influence
of turbulent convection on the general features of
the activity is a more complex question. The second
method compares observational data on the activity
of the Sun and other similar stars. This method has
led to the development of the idea of a single-parameter
gyrochronology relating the activity level to the stellar
age [1].

X-ray and optical observations of numerous late-type
active stars (detected during searches for planets
with large ground-based telescopes and the Kepler Space 
Mission) allowed us to move forward in comparison the activity of stars
of various ages and the Sun. We mention here new 
studies of the relationship between the X ray emission
and rotation [2] and further development of the gyrochronology
based on the comparison between
the chromospheric and coronal activity and 
clarification the role ofmagnetic fields on various scales in
the formation of the activity [3--5].

We are trying to understand what one can say
about the type and level of the Sun’s activity at the
epoch when the solar-type activity was only beginning
to form. It is clear that the activity of some
stars is saturated. In other words, the X-rays from
the coronas of these stars, or more precisely, the
luminosity ratio $L_X/L_{bol}$, reaches $10^{-3}$ and depends
only weakly on the axial rotation rate. This is the
case for stars with rotation periods from 0.3 to 7 days.
The ages of these stars are definitely less than 800 Myrs.
Their coronas are almost completely occupied by hot
regions (with temperatures of about 10 MK), and
the spottedness of their surfaces can reach tens of
percent. The activity of such stars differs considerably
from that of the Sun.

Here, we consider stars that rotate more slowly,
whose coronal activity is below the saturation level.
At the same time, these stars demonstrate fairly high
chromospheric and coronal activity, with no signs of
circumstellar disks or dominance of the dipole field.

Signs of stellar magnetic activity on the lower
part of the main sequence include the formation and
development of complexes of phenomena in various
atmospheric layers: photospheric imhomogeneities,
flares, and coronal holes. The Sun is the best-studied
magnetically active star, and its spots have
been observed over more than 400 yrs, starting from
1610, the epoch of Galileo. We will call the 11-year cycle 
(its length was about 10 yrs in the 20th century) 
the main cycle of solar activity. Its duration
has varied from 7 to 17 yrs at various epochs, and
the amplitude of the maxima vary over a cycle of 
70--100 yrs (the Gleissberg cycle). The most reliable
datae to improve the accuracy of estimates of
sunspot activity and analyses of the 11-year cycle are
the Wolf numbers $W(t)$, which have been recorded
since 1849. Relations between the areas and the
total magnetic fluxes of sunspots have been reliably
determined, as well as regularities in the appearance
of global maxima and minima in the solar activity [6].

The solar activity is undoubtedly connected with
the evolution of magnetic fields. In the past, this
was taken to be associated with sunspots, which
were viewed as hills of magnetic fields. Sunspots
are included in active regions occupying comparatively
small areas of the surface. In addition to the
local magnetic fields of active regions, there also exist
weaker large-scale fields. The dipole magnetic field
of the entire Sun is observed near the solar poles, and
this field changes sign every 11 yrs. This polarity
reversal of the global dipole field occurs near the
cycle maximum. Thus, the configuration of the largescale
magnetic fields is re-established every 22 yrs,
forming a magnetic cycle.

Studies of the cycles of stars in the lower part of
the main sequence were begun relatively recently, in
the middle of the 20th century. The basis of these
studies is the works of Wilson [7, 8],
which were devoted to the analysis of the intensities of
the CaII H and K lines as an indicator of chromospheric
activity; these studies were successfully continued in
the HK project. The analysis of data series obtained over
intervals of more than 30 yrs has shown that $85\%$ of
111 objects studied are variable, with cyclic variations
with durations of about 7 yrs found in $60\%$, linear
trends on scales of $\sim 25\:$yrs or more found in $12\%$, and
irregular variations found in $13\%$ [9]. Relationships
between the cycle characteristics (amplitudes and
periods) and the stellar parameters have been found,
and the activity level has been studied for various
groups of stars with various ages and convection zone
depths.

Projects continuing studies of stellar chromospheric
activity are being carried out at many observatories.
Programs of systematic studies of photospheric
activity were started in the 1990s [10].
Data from the All Sky Automated Survey (ASAS)
and the MOST, CoRoT, and Kepler space
projects are being used for a number of studies. Spot
cycles similar to the 11-year solar cycle have been
found, as well as cycles involving changes in the
active longitudes and the areas and distributions of
spots; these must be taken into account in an analysis of
stellar cycles. Examination of the evolution of stellar
activity for various groups of F--M stars has shown
that dynamo processes occurring at different levels
of the convective zones are responsible for the cycle
formation.

Various solar phenomena are associated with both
large-scale fields (coronal holes and active longitudes)
and local fields (active regions and spots).
Both the large-scale fields and the quasi-biennial
cycle are probably related to dynamo processes occurring
near the base of the convective zone, at the
tachocline at a depth of 0.3 solar radii. On the other
hand, according to helioseismology data, the bases of
sunspots are located only 40--50 thousands km below the
photosphere. This means that local fields are directly
related to phenomena occurring beneath the photosphere.

The evolution of activity on time scales of about
a billion years was studied in [3], and evidence was
found that cycles are formed at a definite stage of a
evolution of activity. Since the type and level of activity
depends on the axial rotation rate of a star,
it is necessary to ascertain the relationship between
characteristics of the cycle and the periods of axial
rotation. The studies [11--13] were devoted to this
problem (see also the review [14]). These studies
repeated earlier analysis, supplementing the HK-Project 
data with more recent observations. However,
the authors note that the samples studied contain
objects of various types, including binary stars. Due
to the insufficient frequent observations and their total 
duration, it has not been possible to reliably estimated the
cycle durations, especially when only a few periodic
variations have been observed for a single star.

The goal of our present study is to clarify the conditions
under which regular cyclic changes in activity
are formed. We focus mainly on estimates of the
stage of the evolution of the activity when the cycle is
formed. First and foremost, we have re-analyzed the
relationship between the cycle duration and the rate of
axial rotation using only reliable data. We have also
tried to identify the physical conditions under which
the chaotic behavior of plasmamotions is transformed
into regular cyclic changes. Such studies can facilitate
the development of dynamo theory and 
our understanding of magnetic activity.

\section{When is the Cycle Formed?}

In addition to data on the long-term variability
of the chromospheric radiation of more than 100 of
the nearest stars studied in the HK project, information
on chromospheric activity has been obtained in
searches for exoplanets in the Northern and Southern
hemispheres. An analysis of these observations can be found
in [3], where it is shown that the index of chromospheric
activity $\log R'_{HK}$ for G stars varies within
narrow limits from --4.9 to --5.1. Note that the level of
the solar chromospheric activity near the cycle maximum,
$\log R'_{HK} = -4.85$, exceeds the level of the main
group of these stars. Younger stars of the Hyades
Cluster were also studied in [3]. Various activity levels
are shown in the histogram in Fig. 1a, where the
second maximum with higher activity includes stars
with ages of about 600 Myrs in the Hyades.

In addition to field stars and open clusters, stars
with high spatial velocities were observed in the far
ultraviolet (FUV: 1350--1780$\;\AA$ ) by the GALEX
(GALaxy Evolution EXplorer) spacecraft [15]. UV
photometry has revealed stars with both high and
low activity levels among the 1360 stars studied.
Figure 1b presents a histogram of the level of chromospheric
activity for these stars. Comparing the
two histograms in Fig. 1, we can follow the evolution
of the chromospheric activity as a function of age (the
three maxima in Fig. 1b for $\log R'_{HK} =$
--4.45, --4.8, and --5.0).

Modern data enable comparisons of the levels of
chromospheric and coronal activity of low-mass stars
of various ages. Figure 2 presents a diagram showing
the main branch of stars connecting stars with low
activity levels and saturated stars. The straight line
connects young stars with ages of hundreds of millions
of years and old stars with ages comparable to
or exceeding the solar age (4.5 Gyrs). This dependence
differs only slightly from the one found earlier
in studies of single-parametric gyrochronology [1]. This
means that the axial rotation is indeed themain factor
determining activity levels.

We analyzed the chromosphere--corona diagram
earlier; we focus here on stars with clearly pronounced
cycles. According to the classification of the HK
project, stars with  Excellent and Good type cyclles are 
concentrated along a straight line in the
chromosphere--corona diagram that is close to the
line for the single-parametric gyrochronology. All these
stars are bounded from above by a certain level of
chromospheric activity. The star with Excellent cycle
V2292 Oph (HD 152391, G7V) is also located in
the same place, with $\log R'_{HK} = -4.4$; this activity
significantly exceeds the levels of the remaining stars
with pronounced cycles.

Most BY Dra-type stars are located in a compact region
in the chromosphere--corona diagram, in the region
$R_X = \log(L_X/L_{bol}) = -4.5$ and 
$\log R'_{HK} = -4.4$.
The activity of late-type stars is characterized by variability
of the optical continuum associated with their axial
rotation. This is usually indicated by the presence of
surface inhomogeneities (mainly cool spots). Stars
with numerous spots are classified as BY Dra variables.
In addition to the rotational modulation, some
of these stars display long-term variations in their
optical continuum, which are sometimes regular or
cyclic. Stars of this type possess fairly high activity at
all altitudes in their atmospheres. For example, the
area occupied by spots can exceed the spotted area at
the solar maximum by a factor of 100.

In general, BY Dra variables form a large class
of objects, including both comparatively young stars
with noticeble lithium abundances and older stars with
low chromospheric activity (Figure 2 presents only
some of these stars, studied in connection with their
lithium abundances). Note the significant number of
binary stars among objects of this type. This binarity
may be related to the fact that these stars maintain
high activity over longer time scales than do similar
single stars. This has been attributed to interactions
between the angular momenta of the orbital and axial 
rotation. Thus, the diagram allows us 
to indicate the region to which stars with cycles extend.
We wish to clarify the epoch in a star’s life when a
regular cycle arises. It is well known that the scale of
stellar ages in the lower part of the main sequence is
based on observations of open clusters. The rotation
of most stars in such clusters decelerate with time,
due to the loss of the angular momentum of the stellar
rotation via the outflow of the magnetized stellar
wind. The ages of these clusters can be taken to be
known from stellar-evolution computations. Young
open clusters are fairly well studied up to the Hyades
Cluster, whose age is about 600 Myrs — closest
to the region of interest,
$R_X = \log(L_X/L_{bol}) = -4.5$ and 
$\log R'_{HK} = -4.4$
in the chromosphere--corona diagram.
We can consider the scale of ages of low-mass
stars to be well determined, from hundreds of millions
to billions of years [16].

\begin{figure}\centering                      
\includegraphics[width=\linewidth]{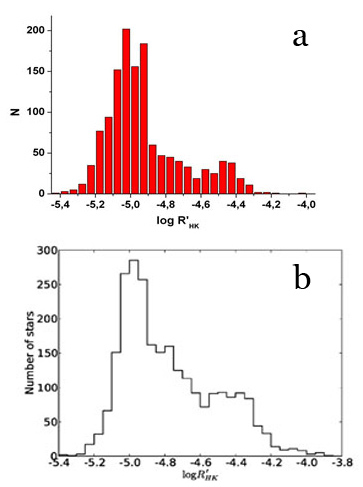}
\caption{Distribution of levels of chromospheric activity 
according to (a) data from searches for exoplanets and (b) GALEX
space observations [15].}
\end{figure}

\begin{figure*}[!t]\centering                       
\includegraphics[width=\linewidth]{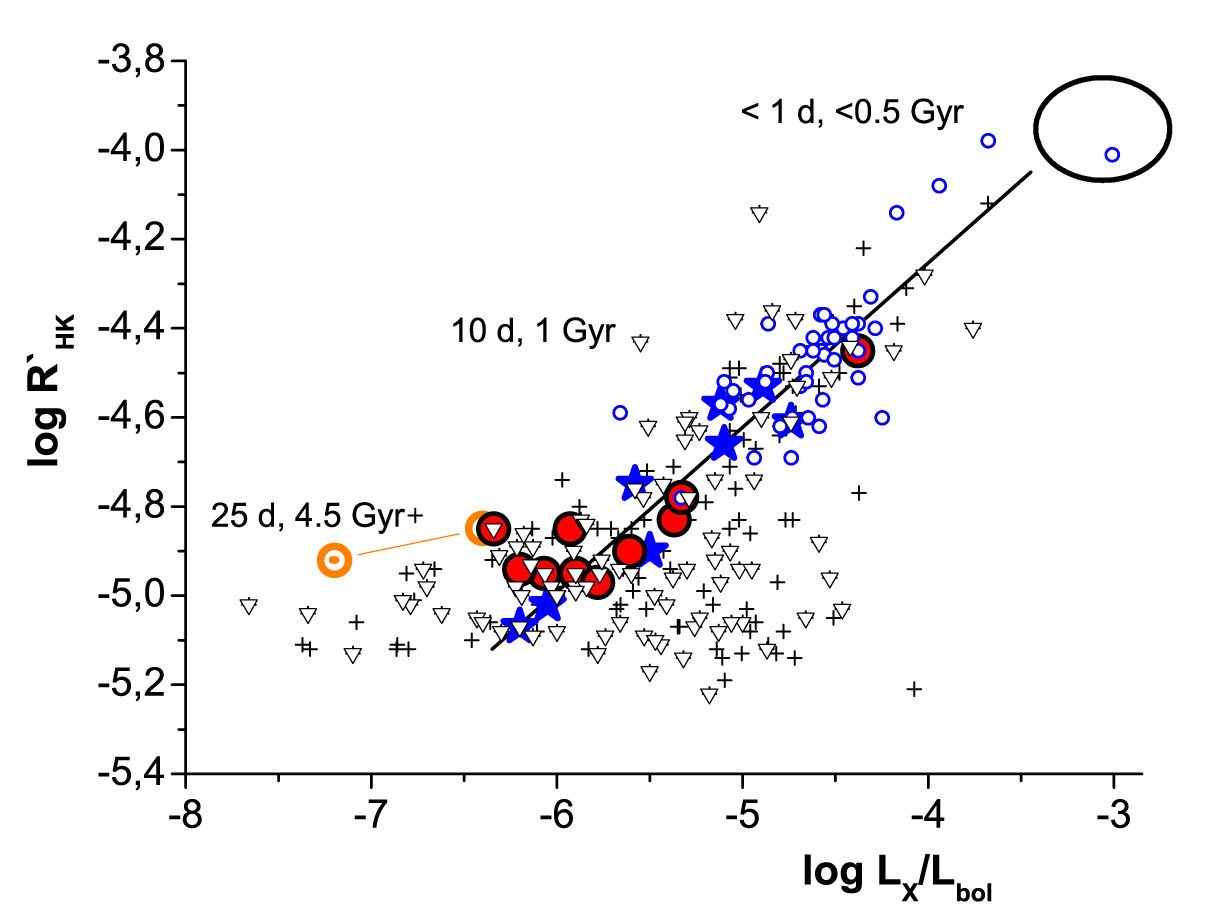}
\caption{
Indices of chromospheric and coronal activity for active late-type stars. Saturated stars are indicated by an ellipse. The
straight line corresponds to the main scenario for the evolution of activity [3], and almost coincides with the line corresponding
to the single-parametric gyrochronology [1]. The stars below the line apparently correspond to another activity-evolution
scenario [3]. Stars with Excellent cycles are shown by red circles,  stars with Good cycles by blue asterisks, BY Dra stars 
by blue circles, stars with high proper motion by triangles, and field stars by crosses. The points for the Sun at its cycle 
maximum and minimum are connected with a line segment. The periods of axial rotation and stellar ages for given levels of 
chromospheric activity are also indicated.}
\end{figure*}

Considerable progress in the development of gyrochronology
was achieved in [17], where an index of
coronal activity is expressed through a combination
of parameters such as the radius and rotation period
($L_X/L_{bol}$ is proportional to $R^\alpha P^\beta$). 
This makes it possible to reduce the spread in the observed ratio
$L_X/L_{bol}$ as a function of the rotation period. Note
that this analysis does not use the Rossby number,
although LX/Lbol clearly depends on this quantity.
We used the physically reasonable solution [17] with
the parameters $\alpha = -4$ and $\beta = -2$, which corresponds
to the proportionality coefficient $k = 1.86 \times
10^{-3}$. The X-ray luminosity for unsaturated regions is
$L_X \sim P^{-2}_{rot}$, in agreement with the conclusions of [18,
19]. However, there are no active stars with the
solar luminosity with rotational periods Psat shorter
than 1.6 days (see formula (10) in [17]). It is known
that this regime with saturated activity corresponds
to $L_X/L_{bol} \sim 10^{-3}$, and depends very weakly on the
rotation period.

There exist some stars with periods from $P_{sat} =
1.6$ days to $P_{sat} = 10$ days whose activity is either
saturated or below the saturation level. Let us note
two characteristic features of stars with similar periods.
First, these stars have lithium abundances only
slightly lower than those of young stars that rotate
with periods of several hours. Second, the activity
of such stars clearly depends on the stellar radius:
changes of $5-10\%$ in the observed radii affect both
the activity and the lithium abundance.

\begin{table*}[!t]\centering
\bigskip \begin{tabular}{|c|c|c|c|c|c|c|} \hline
Star name & Sp & $P_{rot}$, d & S,$\%$ & $L_X$, erg/s & $R_X$ \vbox{\vskip12pt} & 
$\log R'_{HK}$\\
\hline
\strut Active Sun & G2 V & 25 & 0.3 & $10^{27}$ & -7 & -4.90 \vbox{\vskip11pt}\\ \hline
\strut Young Sun & & 10 & 3 & $10^{29}$ & -4.4 & -4.45  \vbox{\vskip11pt}\\ \hline 
\strut BE Cet & G2 V & 8 & 3 & $10^{29}$ & -4.4 & -4.43 \vbox{\vskip11pt}\\ \hline 
\strut $\kappa^1$~Cet & G5 V & 9 & & $10^{29}$ & -4.4 & -4.42 \vbox{\vskip11pt}\\ \hline 
\strut EK Dra & G0 V & 3 & 10-20 & $10^{30}$ & -3 & -4.15 \vbox{\vskip11pt}\\ \hline 
\end{tabular}
\caption{Parameters of G dwarfs of various ages \vbox{\vskip15pt}}%
\end{table*}

Thus, a refinement of the age scale together
with a new analysis of the relationship between
activity levels and rotation periods makes it possible
to determine stellar ages in the interval from
100 Myrs to several billion years. For example,
the selected region in the chromosphere--corona
diagram where $\log(L_X/L_{bol}) = -4.5$ corresponds
to the rotation periods of 10.5 days. This value is
supported by a preliminary analysis of cluster data
obtained by the Kepler spacecraft [16]; in particular,
there is a single relation between the rotation periods
and the masses of late-type stars in NGC 6811, with
an age of one billion years. It is shown that, for solarmass
stars, a rotation period of 10.8 days corresponds
to this age.

We can conclude that solar-type activity is established
in G stars with rotation periods $\ge 10$~days, and,
consequently, with ages $\ge 1$~Gyrs. These phenomena
are well studied for the present-day Sun. Physical
processes resulting in the formation of spots, faculae,
and flares are determined by specific interactions
between large-scale and small-scale solar magnetic
fields. The formation of regular cyclic changes is essential
for the solar activity. Further, we will apply the
concept of "solar-type activity" to stars with periods
of 10 days and longer. We will call a G dwarf with a
rotation period of about 10 days, corresponding to an
age of one billion years, a young Sun. It is believed
that the cycle on such a star has stars have already 
formed.

Solar-type activity differs from processes occurring
on other stars rotating with periods from several
hours to several days and demonstrating saturated
activity. Though these young stars sometimes exhibit
phenomena similar to those observed on the Sun,
the role of the large-scale dipole fields dominates on
these stars; global restructing of the entire corona can
occur, and the character of nonstationary processes
differs strongly from those on the contemporary Sun.
For example, the energy and frequency of superflares
of Gstars differ for stars with periods shorter than and
longer than 10 days [20].

Let us briefly describe the activity of the young
Sun. A significant amount of hot plasma with temperatures
of 5--8MK exists in coronae of stars of such an age.
On the Sun, the maximum of differential
emission measure $DEM(T)$ is about 1--2 MK; on the
young Sun, this maximum is shifted to 6MK, and exceeds
the present solar value near the cycle maximum
by two orders of magnitude. Some parameters of G
dwarfs of various ages are presented in the Table 1.

In the portion of the diagram (see Fig. 2) where
the BY Dra-type stars are concentrated, V2292 Oph (HD
152391, G7V), with a period of 11 days, is the only
star with an Excellent cycle with high chromospheric
and coronal activity. Since there are quite a few spots
on this star, as on BE Cet, the spots determine the
variability of the optical continuum during the cycle,
and there is an anti-correlation between the longterm
variations observed in the continuum and in the
chromospheric CaII H and K lines. It is known that
the Sun and other slowly rotating stars possess 
a correlation between the total radiation of the photosphere
and chromosphere. According to estimates presented
in [21], the young Sun can is undergoing numerous flares
with total energies reaching $10^{34}$~erg, with the occurrence
of superflares being one per 500 yrs. The massloss is 
estimated to be $10^{-11} M_\odot/\textrm{year}$, with
the relative contribution of coronal mass ejections to
the stellar wind being significantly higher than for the
present-day Sun. Some parameters of the magnetic
fields of the young Sun are also discussed in [21].

Finally, we note that cyclic activity is formed on G
and K stars with ages of about 1--2 Gyrs. Starting
from this age, we can speak about  solar-type activity for
low-mass stars.

\section{Relation between Cycle Duration and \\ Rotation Period}

Cycles are distinctive features of solar-type activity.
If a cycle has already formed at an age of 1--2 Gyrs,
the cycle parameters can change in the course of the
star’s evolution. This question has already been studied
using all observed indications of cyclic changes.
However, the use of dissimilar data has hindered the
ability to draw reasonable conclusions. Therefore,
we analyze here only data from the HK project for
stars with Excellent and Good cycles [9], that is,
with reliable rotation periods and cycle durations. We
also added three stars whose cycles were identified
through specially held wavelet analyses [22]. We also refined
the cycle duration of about 14 yrs for HD 149661
using all observations available from 1965 to 2002.
Figure 3 presents the results of our analysis, and
shows a tendency for cycles to become longer with
decelerating rotation for stars rotating more slowly
than the Sun. This suggests that the cycle duration
increases with age. Note that this refers to long
cycles; we do not consider short cycles here due to
the lack of sufficiently reliable data.

It is of interest to investigate this tendency for a
more representative sample of stars. The problem
is that sufficiently uniform and long-term series of
observations are required to establish the cycle durations.
At present, these observations have significant
errors. Direct comparisons of measured cycle durations
and rotation periods for all late-type stars do not
display any obvious trends. Three considerations are
useful for analyzing this question. First, we should
consider long and short cycles separately. If two
cycles are identified for a star, the choice is obvious. In
the remaining cases, we must apply an additional analysis
to determine the group of cycles to which the cycle
duration corresponds. Second, we should use the data
representation proposed in [11, 13] and examine the
relationship between $\log(P_{cyc}/P_{rot})$ and 
$\log(1/P_{rot})$;
i.e., between the number of stellar rotations over the
cycle and the rotation velocity. Third, we should not
analyze all late-type stars together, since the cycles of
M-stars result from a number of different and complex
processes, hindering identication of a general law for
F--K stars with surface convective zones.

We used the information verified to analyze cases
where the cycle durations are reliably established for
G and K stars (including BY Dra-type stars). We selected
60 stars from the observations of [9, 13, 23] for which
only the main cycle, or two cycles, have been determined.
In some cases, the reliability identification of a
cycle has been based on a wavelet analysis similar to
that applied in [22].

Figure 4 presents the results of our analysis. Data
on short cycles are based on the photometric observations
[13]. The correlation coefficients for the long
and short cycles are 0.833 and 0.928, respectively.
Linear fits $Y = A + BX$ for the two types of cycles
yield the coefficients $A = 3.459$ and $B = 0.918$ for
the long cycles and $A = 3.094$ and $B = 0.884$ for the
short cycles.

Here, we have used purposely the same data representation
as that used in [13]. Although the stellar samples
and the methods used to identify the cycles are somewhat
different, our results are close to those obtained
in [14]. In both cases, the two fitted lines are almost
parallel, but our lines are farther from each other than
those in [14] (see Fig. 10 in [14]).

Using explicit instead of statistical relationships
between the cycle durations and the rotation periods,
we find that the relation displayed is fairly weak. Nevertheless,
these two independent analyses indicate a
common character for the relation discussed. Let
us consider the explicit expression for one type of
cycle, for example, for the linear fit $\log P_{cyc} = A +
(1 - B) \log P_{rot}$. It is clear that $B = 1$ separates
two kinds of solutions, in which the cycle duration
either increases or decreases with the deceleration of
the rotation. The currently available data indicate that
it is most likely that the cycle durations increase with
increasing rotation period. This conclusion agrees
with the results obtained for HK project stars with
pronounced cycles presented in Fig. 3.

\begin{figure}\centering                     
\includegraphics[width=\linewidth]{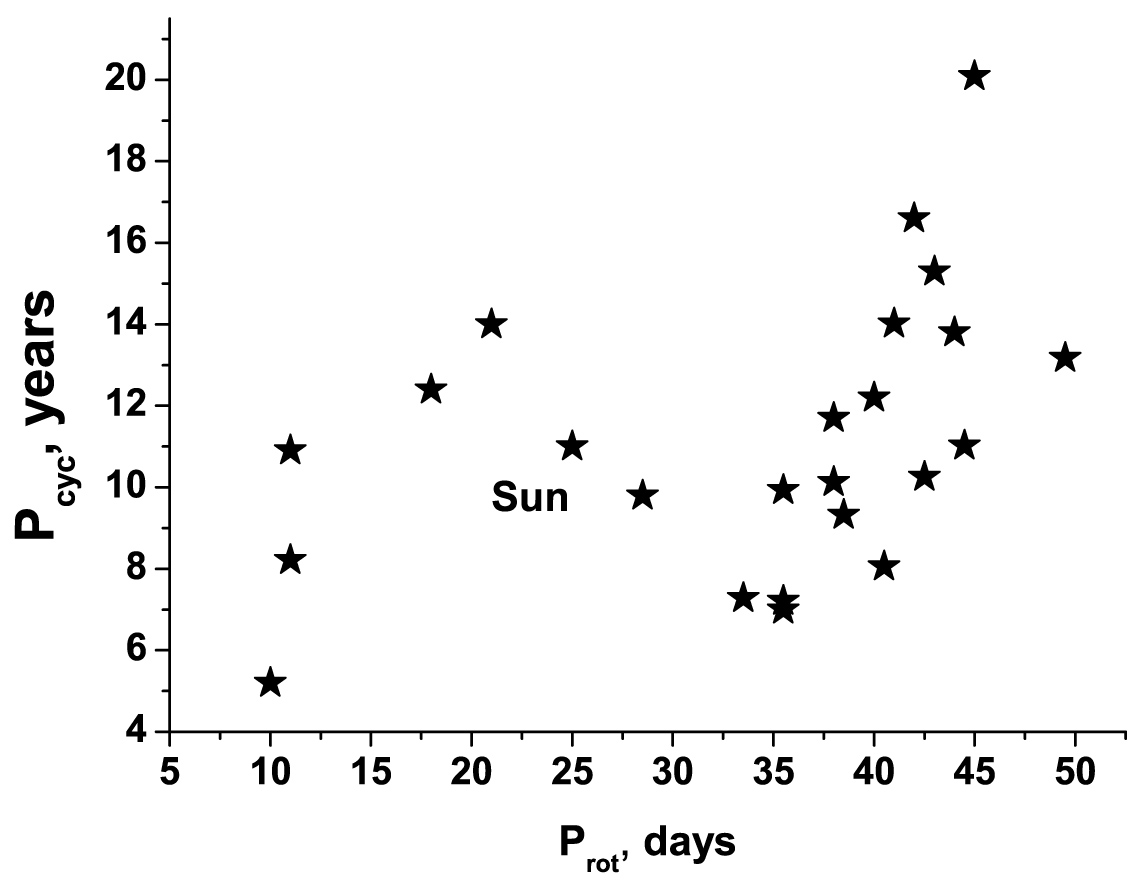}
\caption{
Cycle duration as a function of the rotation period for stars of the HK project.
}
\end{figure}

\begin{figure}\centering                     
\includegraphics[width=\linewidth]{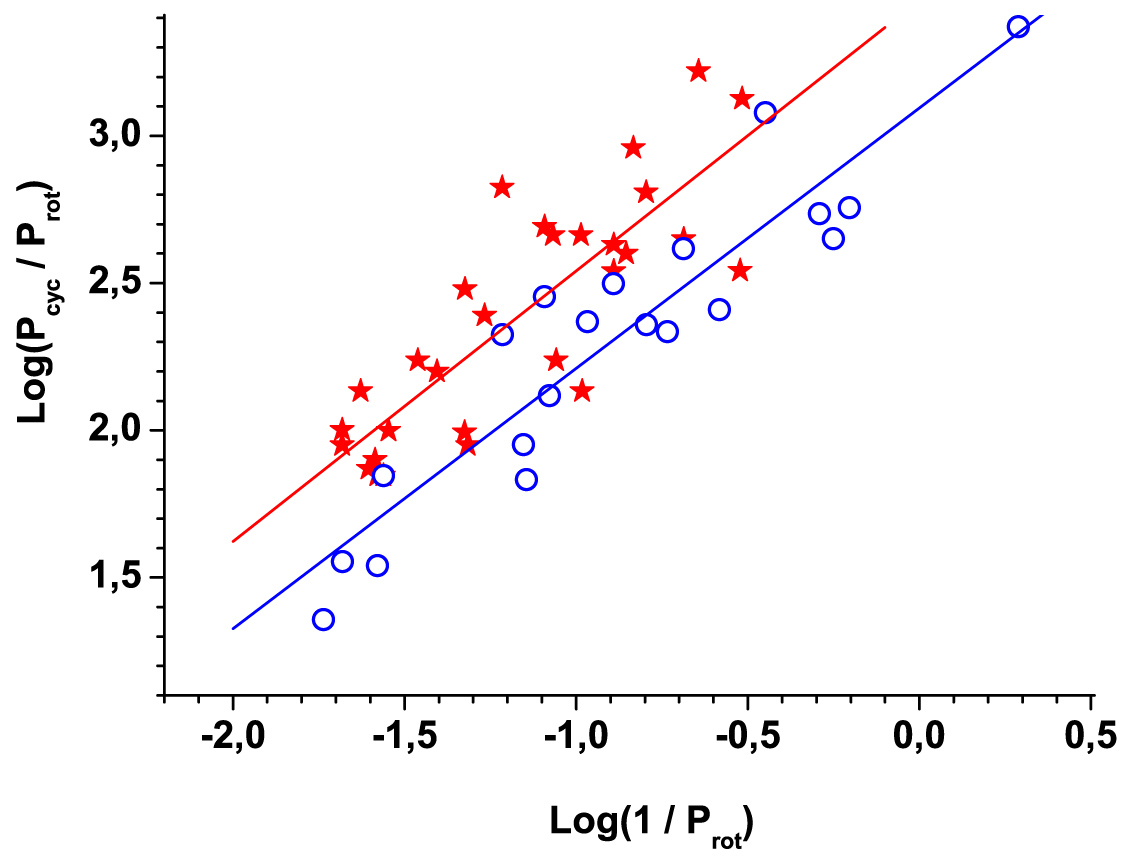}
\caption{
Comparison of cycle durations and rotation periods for long (asterisks)
and short (circles) cycles.
}
\end{figure}

Our analysis enables us to estimate the duration
of the solar cycle for the epoch when this cycle was
formed and become fairly regular. Since the fitted
lines shown in Fig. 4 are almost parallel and most of
the durations $P_{cyc}$ $(N = 49)$ approach the upper line,
we combined the data to obtain a common linear fit
for all the Pcyc values, with the parameters $A = 3.342$
and $B = 0.840$. This corresponds to cycle durations
of 10.2 yrs for the present Sun and 8.7 yrs for the
young Sun.

\section{Discussion. Magnetic Fields on G Stars}

Thus far,  only a few measurements of the magnetic fields 
of low-mass dwarfs were carred out. Similar to 
solar observations, the strength of the field along a
line of sight was directly determined from spectra
containing one or more magneto-sensitive lines, first
and foremost in the visible, and then in the infrared.
Such a study of the Zeeman effect for late-type stars reveal
spots with field strengths of 1--3 kG occupying up
to $10\%$ of the stellar surface. The signal intensities
change with the phase of the rotation period. This
question has been well studied for only a few stars.
Relevant data can be found, for example, in [24, 25].
These data mainly refer to K stars ($\varepsilon$~Eri, 61 Cyg A,
$\sigma$~Dra) and the star $\xi$~Boo A (G8 V). The observed
magnetic fields of more active stars with periods of 6
and 12 days agree with a model in which spots with
field strengths of about 1.5 kG occupy nearly $20\%$
and $10\%$ of the surface, respectively. The magnetic
fields in spots on stars with rotation periods of about
30 days reach 1--2 kG, with the spotted area being
about $2\%$.

\begin{figure}[ht]\centering                   
\includegraphics[width=\linewidth]{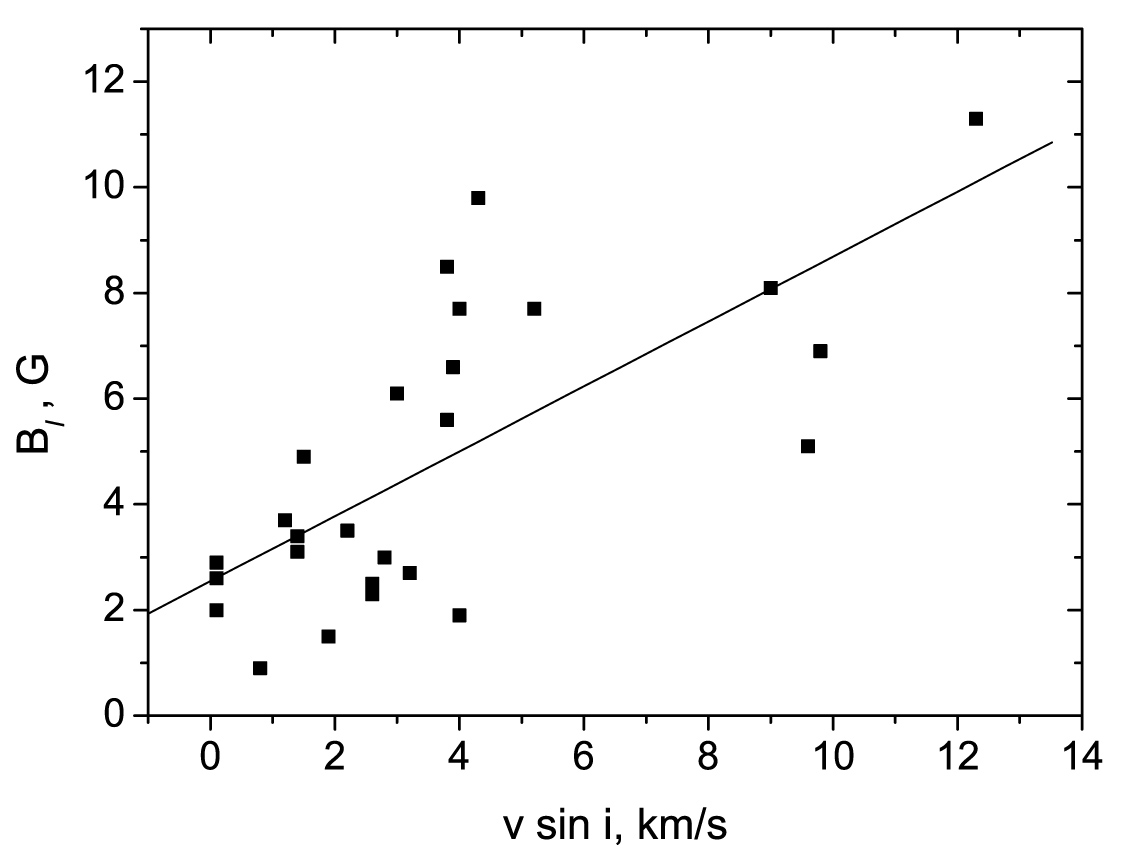}
\caption{
Magnetic fields of main-sequence G stars. 
The straight line shows the linear fit to the data.
}
\end{figure}

These data agree well with current understanding
of stellar magnetic fields. First, variations in
the magnetic-field strength with the phase of the rotation
period, together with spectral data, can be used
tomodel the distribution of inhomogeneities over the stellar
surface. This method of Zeeman-Doppler mapping
enables the separation of the contributions of spots
and large-scale fields, in the first approximation. At
the same time, the main regularities, such as the
relationships between the mean surface field and the
rotation velocity or the total level of stellar activity, are
preserved [26].

Thus, we conclude that the general behavior of
the magnetic fields of G and K stars is similar to
that observed for the present-day Sun: namely, largescale
fields are most distinctly observed in the polar
zones and local fields at lower latitudes. The contribution
of magnetic fields of various scales to the
mean field varies over several years, and depends on
the orientations of the rotation axis and the dipole axis
relative to the line of sight.

Second, new information on the magnetic fields
of solar-type stars has recently been obtained in the
BCool Collaboration program [27], in which large
telescopes carry out spectropolarimetric observations
of 170 late-type stars. From 5000 to 11 000 spectral
lines are simultaneously used to measure the magnetic
field, making it possible to measure the lineof-
sight components of stellar magnetic fields fairly
accurately. The quantity $B_l$ is the signal averaged
over all observations of a star. These spectral observations
can be used to determine both fundamental
parameters of the stars and the levels of chromospheric
activity. Magnetic fields have been detected
in $40\%$ of these stars. The mean line-of-sight field for
the K stars is $|B_l| = 5.7$~G [27], which is higher than
the fields of both G and F stars (3.3 G).

We have briefly analyzed only the data for G
dwarfs, in order to estimate the magnetic-field
strength on the young Sun. The technique used has
enabled us to detect magnetic fields corresponding
to epochs when the global solar dipole reaches its
maxima of about 1 G (for example, this was the case of the Sun 
in 1991 [28]). We took all G stars with $B_l$ exceeding
$3\sigma$ from Table 3 of [27]. We omitted some rapidly
rotating, and thereby some young stars with stronger
magnetic fields. The final list includes 28 G stars
with periods exceeding 7 days (from approximate
estimates based on data about chromospheric indices
taken from Table 5 of [27]). Figure 5 presents our results.
The field magnitudes $|B_l|$ of two stars, $\xi$ Boo A
and 61 UMa, considerably exceed the fields typical
for stars with such rotation rates. This probably
results from uncertainties in the inclination between
the stellar rotation axis and the line of sight [27]. With
these two stars omitted, the correlation coefficient
between $|B_l|$ and $v \sin i$ exceeds 0.70.

Thus, if we omit starswhose fields are not detected
reliably and some stars with very strong fields, we
find the mean field $|B_l|$ reaches 4.72$\pm$0.53~G for G
stars. The relationship shown in Fig. 5 describes
the weakening the mean field as the stellar rotation
decelerates, i.e., as the star ages. The mean field we
have found corresponds to a rotation speed of 4 km/s,
which is twice the value for the present-day Sun. We
can thus adopt this field strength as the mean field
of G stars with rotation periods of about 10 days,
corresponding to ages of 1--2 Gyrs.

Let us compare this value with data on the total
magnetic field of the Sun as a star. The fields
averaged over a Carrington rotation do not exceed
0.5 G at solar maxima. Thus, the mean magnetic
fields of young G stars exceed the fields of present
solar maxima by at least an order of magnitude. Note
that our estimate is confirmed by $|B_l|$ estimates for
stars whose parameters are closest to those of the
young Sun: 8.5 G for V 2292 Oph and 7.7 G for
$\kappa^1$~Cet (according to different data, the field of this
star is $|B_l| = 7.0$~G [29]). This agrees with our general
conclusion above.

Information on the large-scale magnetic fields of
late-type stars is now becoming available. In particular,
slowly rotating stars possess fields with structures
similar to those of solar fields, which change regularly
in the course of the 22-year magnetic cycle. In other
words, a global dipole is almost always present; only
during polarity reversals does the magnetic equator
move in longitude in a way that can be treated like
the manifestation of a large-scale toroidal component.
Active regions are located fairly chaotically, active
longitudes are poorly observed, and toroidal fields are
characterized by comparatively small scales ( they
are local fields). At the same time, G stars with
rotation periods below 12 days display clear largescale
toroidal field components [26]. Though these
toroidal fields dominate over the large-scale poloidal
fields, the difference in their strengths is small. It is
difficult to relate these new finding with spot data for
rapidly rotating G stars.

\section{Conclusion}

We considered the evolution of solar-type activity
in earlier studies. It was noted that 
stars with cycles are located nearly along a straight line in
the chromosphere--corona diagram. Their location
almost along the line yielded by single-parametric
gyrochronology is simply related to the difference in
the ages of stars with Excellent and Good cycles.
However, these stars do not occupy the entire
line up to stars with saturated activity, and reach
only a certain point, with V2292 Oph (HD 152391,
G7V, with an Excellent cycle) and a rotation period
of 11 days located slightly below ths point (BY Dra-type
stars are also concentrated there). There are reasons
to suppose that this activity level corresponds to the
start of solar-type activity with cycles. Using the
new relation between coronal activity and rotation
rate [17] and new data on the ages of open clusters,
we have estimated the age of the young Sun at the
epoch of cycle formation. We have briefly discussed
the activity of this young Sun, whose age is slightly
older than 1 Gyrs. We have used reliable data to
compare the cycle durations with the rotation periods,
and conclude that the cycles lengthen as the stellar
rotation slows. The relationships revealed lead to
estimated cycle durations of 10.2 yrs for the present
Sun and 8.7 yrs for the young Sun, assuming that the
type of activity has not appreciably changed since the
epoch when the cycle was established.

It is clear that the solar activity is due to interactions
between magnetic fields of different scales.
What was the solar activity at the epoch when the
quasistationary development of these processes had
just begun? Data on magnetic fields and the activity
occurring in various atmospheric layers of mainsequence
G stars provide answers to this question.
The discussion of new data on the magnetic fields of
comparatively young G stars presented here can be
helpful for identifying the conditions required for the
cycle formation. Note also that observational facts
can also be used to refine the dynamo theory.

\phantomsection

\section*{Acknowledgments}
\addcontentsline{toc}{section}{Acknowledgments} 

This work was supported by the Russian Foundation
for Basic Research (project 15--02--06271).

\phantomsection
\bibliographystyle{unsrt}


\end{document}